\documentclass[twocolumn,superscriptaddress,nofootinbib,aps]{revtex4}%
\usepackage{graphicx}
\usepackage{times}
\usepackage{textcomp}
\usepackage{amsmath}
\usepackage{amssymb}
\usepackage{bm}
\usepackage{units}
\usepackage{stmaryrd}
\usepackage[compact]{titlesec}
\usepackage{xcolor}
\titlespacing{\section}{0pt}{*0}{*0}
\titlespacing{\subsection}{0pt}{*0}{*0}
\titlespacing{\subsubsection}{0pt}{*0}{*0}

\renewcommand{\vec}[1]{\boldsymbol{#1}}

\begin{document}
\preprint{0}
\renewcommand{\thefootnote}{\roman{footnote}}

\title{Gold-induced nanowires on the Ge(100) surface yield a 2D, and not a 1D electronic structure}

\author{N. de Jong}
\email{n.dejong@uva.nl} 
\address{Van der Waals-Zeeman Institute, Institute of Physics (IoP), University of Amsterdam, Science Park 904, 1098 XH, Amsterdam, The Netherlands}

\author{R. Heimbuch}
\address{Physics of Interfaces and Nanomaterials, University of Twente, 7500 AE Enschede, The Netherlands}

\author{S. Eli\"ens}
\address{Institute for Theoretical Physics, Institute of Physics (IoP), University of Amsterdam, Science Park 904, 1098 XH, Amsterdam, The Netherlands}

\author{S. Smit}
\address{Van der Waals-Zeeman Institute, Institute of Physics (IoP), University of Amsterdam, Science Park 904, 1098 XH, Amsterdam, The Netherlands}

\author{E. Frantzeskakis}
\address{Van der Waals-Zeeman Institute, Institute of Physics (IoP), University of Amsterdam, Science Park 904, 1098 XH, Amsterdam, The Netherlands}

\author{J.-S. Caux}
\address{Institute for Theoretical Physics, Institute of Physics (IoP), University of Amsterdam, Science Park 904, 1098 XH, Amsterdam, The Netherlands}

\author{H.J.W. Zandvliet}
\address{Physics of Interfaces and Nanomaterials, University of Twente, 7500 AE Enschede, The Netherlands}

\author{M. S. Golden}
\email{m.s.golden@uva.nl} 
\address{Van der Waals-Zeeman Institute, Institute of Physics (IoP), University of Amsterdam, Science Park 904, 1098 XH, Amsterdam, The Netherlands}

\begin{abstract}
Atomic nanowires on semiconductor surfaces induced by the adsorption of metallic atoms have attracted a lot of attention as possible hosts of the elusive, one-dimensional Tomonaga-Luttinger liquid. The Au/Ge(100) system in particular is the subject of controversy as to whether the Au-induced nanowires do indeed host exotic, 1D metallic states. In light of this debate, we report here a thorough study of the electronic properties of high quality  nanowires formed at the Au/Ge(100) surface. The high resolution ARPES data show the low-lying Au-induced electronic states to possess a  dispersion relation that depends on two orthogonal directions in k-space. Comparison of the E($k_x$,$k_y$) surface measured using high-resolution ARPES to tight-binding calculations yields hopping parameters in the two different directions that differ by a factor of two. Additionally, by pinpointing the Au-induced surface states in the first, second and third surface Brillouin zones, and analysing their periodicity in $k_{||}$, the nanowire propagation direction seen clearly in STM can be imported into the ARPES data. We find that the larger of the two hopping parameters corresponds, in fact, to the direction perpendicular to the nanowires ($t_{perp}$). This, the topology of the $E$=$E_F$ contour in $k_{||}$, and the fact that $t_{||}$/$t_{perp}$$\sim$0.5 proves that the Au-induced electron pockets possess a two-dimensional, closed Fermi surface, and this firmly places the Au/Ge(100) nanowire system outside potential hosts of a Tomonaga-Luttinger liquid. We combine these ARPES data with scanning tunneling spectroscopic measurements of the spatially-resolved electronic structure and find that the spatially straight - wire-like - conduction channels observed up to energies of order one electron volt below the Fermi level do not originate from the Au-induced states seen in the ARPES data. The former are rather more likely to be associated with bulk Ge states that are localized to the subsurface region. Despite our proof of the 2D nature of the Au-induced nanowire and sub-surface Ge-related states, an anomalous suppression of the density of states at the Fermi level is observed in both the STS and ARPES data, and this phenomenon is discussed in the light of the effects of disorder.

\end{abstract}

\maketitle

\section*{Introduction}
Shortly after the seminal works of Tomonaga and Luttinger \cite{tomonaga50ptp,luttinger63jmp}, it was
understood that one-dimensional electron gases (1DEG) display many interesting non-Fermi liquid properties \cite{1965_Mattis_JMP_6,1967_Theumann_JMP_8,1974_Dzyaloshinskii_JETP_38,1974_Mattis_JMP_15,1974_Luther_PRB_9,1974_Everts_SSC_15,1975_Heidenreich_PLA_54,1975_Efetov_JETP_42}.
Their universal validity was first fully appreciated in the 1980\textquotesingle s, in
particular by Haldane who coined the term Luttinger liquid
\cite{1981_Haldane_JPC_14,1981_Haldane_PRL_47}.  In contrast to higher-dimensional electron gases, the spectral properties of 1DEG cannot be understood in terms of electron-like quasiparticles
but rather in terms of bosonic collective spin and charge modes \footnote{Recent developments extending TLL theory \cite{2007_Khodas_PRB_76,2010_Schmidt_PRL_104,2010_Schmidt_PRB_82,2010_Pereira_PRB_82,2012_Imambekov_RMP_84} reinstate a fermionic quasi-particles picture in terms of spinons and holons, in a sense reconnecting the one-dimensional case with Fermi liquid theory, but this does not change the low-energy predictions to be discussed in this paper.}. The remarkable conclusion that
these degrees of freedom can therefore propagate separately with different
velocities, known as spin-charge separation, is one of the hallmarks of a Tomonaga-Luttinger liquid
(TLL). Correlations furthermore display characteristic power-law behavior with certain
universally-related exponents determined by a single interaction
parameter  \cite{voit95rpp,GiamarchiBOOK}.

Despite its theoretical appeal and ostensible universality, finding unambiguous realizations of TLLs has proved challenging. In particular, the great desire to `see' Luttinger liquids must be tempered by the hard requirement that simpler explanations do not exist (Occam's razor). Up to now, carbon nanotubes
 \cite{ishii03nat,lee04prl}, organic crystals with highly anisotropic
 bulk properties \cite{denlinger99prl,hager05prl,claessen02prl}, and GaAs
 channels \cite{Auslaender02Sc} are the most credible examples of classes of
 materials able to display the exotic effects associated to TLLs. Recently, self-assembled atomic nanowires on semiconductor
 surfaces have attracted a lot of attention as further candidates. These systems appear to
 offer the perfect playground to study the electronic properties of
 the smallest conceivable conducting channels. More specifically the
 system of Au-induced nanowires on the Ge(001) surface has been introduced, in which nanowire-like objects that can be 100's of nm long appear clearly in scanning tunneling microscopy (STM) topographic images with an inter-nanowire separation of 1.6 nm \cite{wang04prb}. The exact structure of these nanowires is still a subject of debate \cite{wang04prb,wang05ss,houselt08prb,schafer08prl,mocking10ss,lopez10prb,meyer11prb,blumenstein13jpcm}. LEED measurements reveal a basic c(8$\times$2) periodicity, with an additional periodicity seen on top of the nanowires - referred to as the \textquotesingle VW\textquotesingle structure - results in a (8$\times$4) reconstruction \cite{blumenstein13jpcm,lichtenstein15ss,park14prb}. 
One structural model that fits most of the experimental data is the giant missing row reconstruction model \cite{houselt08prb}. This picture naturally explains that the depth of the troughs between wires to be larger than a single layer of atoms \cite{kockmann09jpcc}, and also rationalizes the difference between the occupied and unoccupied topographic images measured in STM \cite{sauer10prb}. In addition, the fact that the local density of states (LDOS) observed in the troughs is larger than that on the wires themselves \cite{heimbuch12natphy}, and the increased surface corrugation
observed in SPA-LEED measurements \cite{lichtenstein15ss} can also be explained in the giant missing row model, in which the top of the nanowires is formed by Ge-Ge dimers, with the troughs consisting of Ge(111) facets covered in Au trimers \cite{houselt08prb}.  Density functional theory-based calculations predict that the most simple version of this model is not energetically favorable \cite{sauer10prb}, and in its basic form it also does not contain the VW superstructure that is observed with both STM and LEED \cite{blumenstein11prl,blumenstein13jpcm,park14prb,lichtenstein15ss}. Taken together, these considerations point towards the possibility that the real structure for Au/Ge(100) is a more complicated version of the giant missing row model in which the Au atoms are incorporated into the germanium structure \cite{sauer10prb}. In any case, it is beyond debate that the Au/Ge(100) nanowire structures seen in STM are not simply chains of Au adatoms lying on top of the germanium surface, but are in fact complicated 3D surface reconstructions associated with a large increase in vertical corrugation. 
 
Experimental observations interpreted as indicating TLL-like behavior in the Au/Ge(100) system have been subject of controversy in the literature.
In STS-based measurements of the LDOS, a TLL system should show its face as a dip in the differential conductivity around zero-bias and a characteristic power-law scaling behavior for the LDOS around the Fermi level which exhibits a universal dependence on the temperature and energy away from E$_F$ \cite{voit95rpp, GiamarchiBOOK}. The exponent of the power law, $\alpha$, is a measure of the interaction strength between the electrons \cite{matveev93prl} and should show clearly different values for TLL systems when approaching the end of the 1D chains \cite{kane92prl,schonhammer93jesrp}.
On the one hand, the expected kind of power-law scaling of the density of states has indeed been reported in both scanning tunneling microscopy/spectroscopy (STM/STS) \cite{blumenstein11natphy} and angle resolved photo-emission spectroscopy (ARPES) data \cite{meyer11prb,meyer14prb}.
Furthermore, straight features in constant energy $E(k_x,k_y)$ maps in ARPES \cite{meyer11prb,meyer14prb} and linear conduction
pathways observed in the troughs between the nanowires in STM data \cite{schafer08prl,blumenstein11natphy,heimbuch12natphy} would also seem to point towards the generation of Au-induced electronic states that show significant dispersion only in one $k$-direction, heralding the elusive Tomonaga-Luttinger liquid. On the other hand, this conclusion has been disputed based on results of fully analogous experiments carried out by other groups on the same Au/Ge(100) system \cite{nakatsuji09prb,houselt09prl,nakatsuji11prb,niikura11prb,nakatsuji12natphy,park14prb}. The data are argued to be more consistent with the Au-induced surface states being two-dimensional in nature \cite{nakatsuji09prb,park14prb}. Thus, experimentally speaking, it is fair to say that the situation appears undecided. 

From a theory point of view, we note that the observation of a TLL in a solid state system like that of Au/Ge(001) nanowires raises questions concerning instabilities
of the Luttinger liquid state.  One-dimensional systems are  particularly susceptible to disorder and localization
effects. One can show quite generally that for repulsive interactions, Gaussian (random) disorder is a relevant perturbation leading to a pinned charge density wave or random antiferromagnetic phase at low energies \cite{Giamarchi88prb,GiamarchiBOOK}.
A quasi-periodic disorder potential can lead to a Mott-like metal-insulator transition, even at
incommensurate fillings \cite{1993_Kolomeisky_PRB_47,1999_Vidal_PRL_83,2001_Hida_PRL_86}.  
Higher dimensional coupling between nanowires or with the substrate also form a major threat to the observation of TLL physics, 
as such couplings will in general cause a transition to an ordered state or a dimensional crossover below the energy scale set by these couplings. This too often forgotten fact means that TLL physics should not be looked for at asymptotically low energies, but rather at energies {\it above} the scale set by these TLL-destroying couplings, and this search is meaningful only if the effective 1D couplings lie at yet higher energies. Following this line of reasoning, there are also constraints with regards to the regimes of energy and temperature for which TLL behavior could be expected due to the finite length of the nanowires caused by impurities, missing atoms and the finite size of the Ge domains hosting the atomic chains.

In this paper we present a complete study of the Au/Ge(001) system using LEED, ARPES, STM/STS and theory. This combined approach provides a detailed and conclusive picture of the electronic states at and in the vicinity of the Fermi level. The LEED data and STM-imaging attest to the high quality of the nanowire samples. High resolution ARPES data recorded from the same samples enable a robust link to be made between the nanowire propagation direction from STM and the measured dispersion relation of the Au-induced states from photoemission. The ARPES measurements for k-directions perpendicular and parallel to the nanowire show irrefutably that the Au-induced surface states are --- in fact --- 2D in nature, with simple modelling of the E($k_x,k_y$) landscape yielding a $t_{||}$/$t_{perp}$ ratio of $\sim$0.5. Our STS experiments provide access to sample areas containing only a single domain of Au-induced nanowires, and the LDOS data recorded from individually-addressed troughs and from atop single nanowires have been carefully and thoroughly analysed with respect to the expectations for the case of TLL behavior. The tunnelling data force us to arrive at a conclusion that echoes that from the ARPES investigations, namely that the observed straight conduction channels seen in STS experiments of our high-quality Au/Ge(100) nanowire samples do not show TLL characteristics. \\

\section{Experimental details}
\subsection{Sample preparation}
The Ge(100) substrates were cut from nominally flat, single-side-polished, \textit{n}-type (25 $\Omega$cm) wafers and mounted on molybdenum sample-holders. Contact of the substrates with any other material has been carefully avoided during both preparation and the experiments, and they were cleaned using cycles of prolonged 500 eV Ar$^+$ ion sputtering combined with annealing (via resistive heating) at 1100 ($\pm$ 25) K. The result were atomically-clean Ge(100) samples, which exhibited a well-ordered p(2$\times$2)/c(4$\times$2) domain LEED pattern \cite{zandvliet98prb,zandvliet03pr}. Subsequently, gold was evaporated onto the clean Ge(100) substrates from a resistively-heated tungsten wire wrapped with high-purity Au (99.995\%). The Au/Ge(100) sample was then annealed at 650 ($\pm$ 25) K for 2 minutes and subsequently cooled down to room temperature by radiation quenching. At no time during the whole Au-evaporation / annealing cycle did the pressure exceed 5 $\times$ 10$^{-10}$ mbar.\\

\begin{figure*}
	\centering 
	\includegraphics[width = 18 cm]{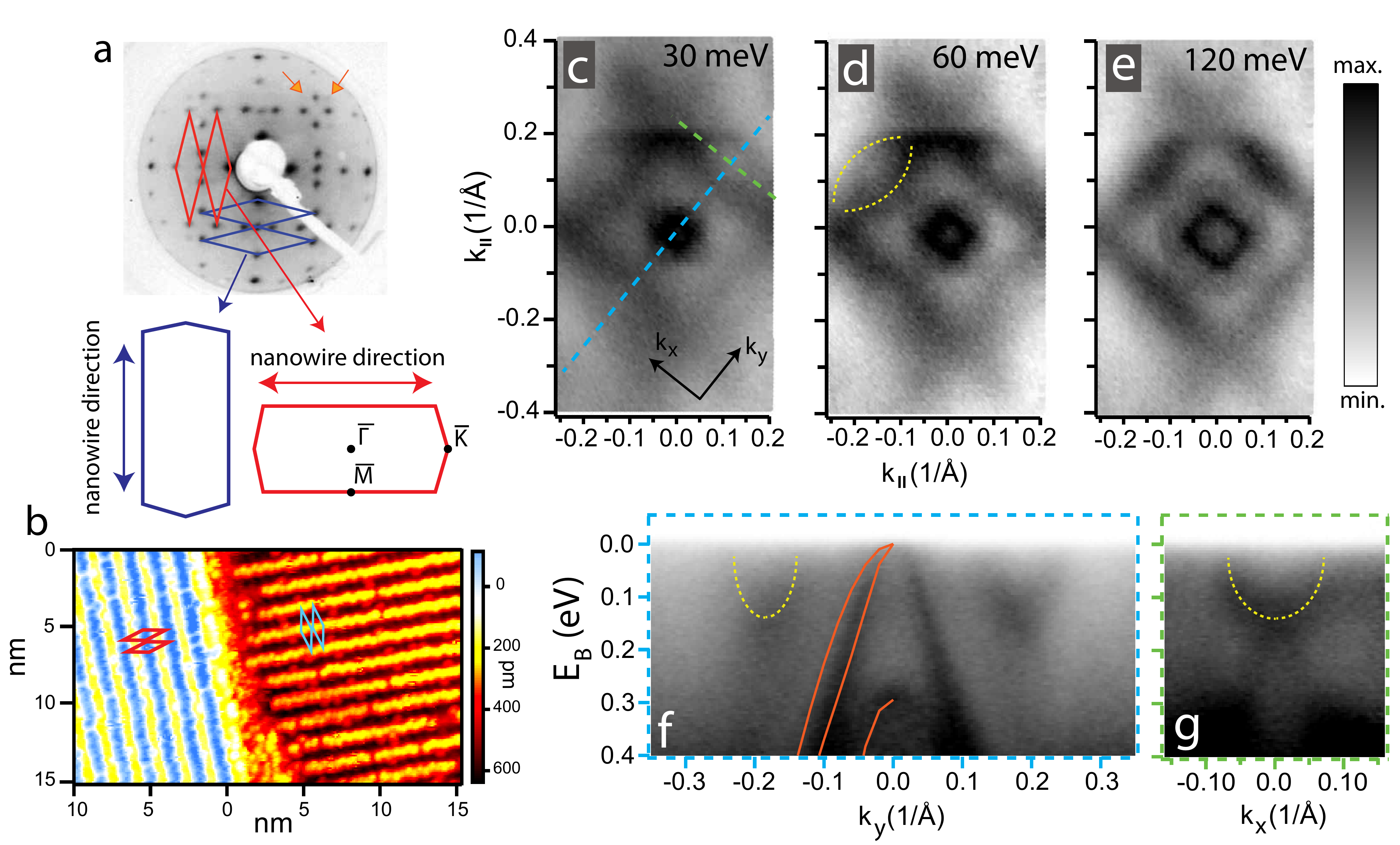}
	\caption{{\bf (a) Representative LEED image of the nanowire sample, recorded with an electron beam energy of 23 eV, showing a clear, dual-domain c(8$\times$2) reconstruction. Two reciprocal surface unit cells are shown superimposed in blue and red, and their respective first Brillouin zones are indicated below the LEED image. (b) STM topograph of  Au-induced nanowires on Ge(100), recorded at room temperature. The unit cells from LEED data are superimposed in red and light blue. (c-e) Constant energy maps --- I(k$_x$,k$_y$) --- measured using ARPES for binding energies, $E_B$ of (c) 30 meV, (d) 60 meV and (e) 120 meV. (f) I(k$_{y}$,E) image, which is a cut along the light blue dashed line in panel (c). The orange lines in panel (f) superpose the results of density functional theory calculations for bulk Ge from Ref. \cite{jancu98prb}. (g) I(k$_x$,E) image along the green dashed line in (c). The yellow dotted lines in panels (d), (f) and (g) highlight low-lying Au-induced electronic states. All ARPES data was taken with a photon energy of 21.2 eV at a temperature of 20 K.
		}}
		\label{fig:Fig1}
	\end{figure*}

\subsection{ARPES}
ARPES measurements were performed using a lab-based ARPES spectrometer at the University of Amsterdam. This system is equipped with a Scienta SES2002 hemispherical electron analyzer, a monochromatized, high-intensity helium discharge source and a six-axis cryogenic sample manipulator. All ARPES measurements presented here were obtained using a photon energy of 21.2 eV, which corresponds to the HeI$\alpha$ line. The total energy resolution was set to 20 meV, and the angular resolution was 0.2\textdegree, resulting in 0.0085 \AA$^{-1}$ resolution in momentum space. The pressure during the measurements was $<1.0\times 10^{-10}$ mbar and all ARPES data presented here were recorded at sample temperatures between 16 and 20 K. The structural quality of the nanowire samples made in the Amsterdam labs was also checked using a commercial UHV, room temperature STM (Omicron). The topographic results agreed very well with those from the sample preparation runs carried out in the LT-STM system of the University of Twente\textquotesingle s MESA+ laboratory, in which members of the Amsterdam ARPES team were also involved.\\

\subsection{STM}\label{sec:LT-STM}
Cryogenic STM and STS experiments were performed using an Omicron UHV, low temperature scanning tunnelling microscope located at the University of Twente. The differential conductivity was measured by applying a small sinusoidal modulation (amplitude 10 mV and frequency 3985 Hz) to the sample bias. The dI/dV signal was extracted using a SRS830 lock-in amplifier (Stanford Research Systems). The measurements were performed at a base temperature of 4.7 K. Both the sample and the STM scanner are located in the cryostat, so as to minimise thermal drift.\\

\section{Results}
\subsection{ARPES data}\label{sec:ARPESdata}

In Fig. 1(a) and 1(b), LEED and STM data are presented which prove the high quality of the Au-induced nanowire samples. Fig 1(a). show LEED data representative of the measured samples (in this case those prepared in the ARPES system). As indicated by the red and blue diamond shapes, the mm-sized LEED beam picks up a c(8$\times$2) and c(2$\times$8) pattern, indicating a dual-domain morphology, echoing that seen in the nanoscale STM topograph shown in Fig. 1b. Additional LEED spots --- highlighted with orange arrows --- are also visible in the image, and are the result of the additional reconstruction on top of the c(8$\times$2) periodicity, attributed to inter-chain interactions \cite{blumenstein13jpcm,lichtenstein15ss}. The two surface Brillouin zones (SBZ) associated with these two nanowire domains are sketched below the LEED image, together with an indication of the appropriate nanowire propagation direction in each case.  

Fig. 1(b) shows the typical topographic signature of long, evenly-spaced lines which has made Au-nanowires on Ge(001) such a appealing model system for possible TLL behavior. As the superimposed (red/light-blue) unit-cell schematics indicate, both nanowire domains show c(8$\times$2) surface unit-cell periodicity, which results from the 90 degree rotation of the Ge dimer rows at each of the single unit-cell step edges of the underlying Ge(001) substrate \cite{schafer08prl}. STM investigations over larger fields of view (see, for example, Fig. \ref{fig:dIdVcurves}(a)), indicate a low density of defects in such nanowire samples, and also confirm the high nanowire coverage over the sample surface.

Figs. 1(c) to (e) display ARPES constant energy maps showing the k$_x$ and k$_y$-dependence of the low-lying electronic states of the Au/Ge(001) nanowire system at the binding energies, E$_B$, indicated.
Figs. 1(f) and (g) present the dispersion relation of the key states along the directions in k-space highlighted in panel (c). Two distinct types of bands are observed. In panel (f) parabolic, hole-like bands centered on $\Gamma$ are seen, which are bulk-derived germanium states\footnote{We refer to these states as bulk-derived states, rather than bulk states since the data presented in Appendix\ref{sec:kzdep} and in Ref. \onlinecite{meyer14prb} show these hole-like, Ge-related bands to show no significant dispersion in k$_{z}$. This is due to the fact that the tops of the valence bands are perturbed by the Au-Ge interface at the surface, so as to form sub-surface 2D states, see Appendix\ref{sec:kzdep} for more details.}, illustrated by the excellent agreement with the orange lines tracing the corresponding states in band structure calculations based on Ref. \onlinecite{jancu98prb}.
Panels (c)-(e) show that the second type of bands compose four electron pockets arranged in a square-like pattern around the $\Gamma$-centered hole-like features.
In terms of the lively, TLL-inspired discussion of ARPES data on Au/Ge(001), it are the four electron pockets in Fig. \ref{fig:Fig1}(c-e) that play the central role \cite{schafer08prl,nakatsuji09prb,meyer11prb,nakatsuji11prb,blumenstein13jpcm,meyer14prb}.
ARPES data from measurements taken at the synchrotron presented in Appendix\ref{sec:kzdep} show that these electron pockets do not show significant k$_z$ dispersion, as expected for states located at the surface that are induced by the Au decoration of the Ge(100). 

That these states originate from the modification of the Ge(001) surface electronic and geometric structure as a result of the deposition of the gold atoms is more or less the only topic of consensus in the literature with respect to the electron pockets observed in ARPES, and there is certainly no agreement as to whether these states support a 1D or 2D-like dispersion relation.
In common with all published ARPES data \cite{schafer08prl,nakatsuji09prb,meyer11prb,nakatsuji11prb,meyer14prb}, the photon beam-spot used in our experiments is significantly larger than the average, nanometric size of the nanowire domains.
This means that the measured ARPES signal is inescapably the sum of the signal from collections of the two orthogonal nanowire domains.
Therefore, as in previous studies \cite{meyer11prb,meyer14prb}, we conclude that the I(k$_x$,k$_y$) ARPES images represent, in fact, the sum of two, pairs of orthogonal nanowire-related features, and that the final result is thus a total of four features with a separation of about 0.2 \AA$^{-1}$ from $\Gamma$: each opposing pair belonging to a single nanowire domain. 

The constant energy maps of Figs. 1(c) to (e) show raw ARPES data, and the signal quality in the first Brillouin zone is certainly sufficient to
make the use of second differentials of the data unnecessary.
What these data show is that, although the electron pockets are anisotropic, they do display significant dispersion in \emph{both} k-space directions, as indicated by the yellow dotted guide-lines in Fig. 1(d).
A degree of curvature was reported for these states previously, but was interpreted to be a small enough perturbation on a tramline-like set of parallel lines so as not to destroy the quasi-1D behavior \cite{meyer14prb}.
Setting aside for the moment possible theoretical considerations as to what extent dispersion perpendicular to the nanowires is permissible for the observation of TLL behavior, it is imperative that high-quality, raw ARPES data are used to examine the dispersion relation for the electron pockets in more detail, as is done in Figs. 1(f) and 1(g). Here, two raw I(k,E) images are shown, taken parallel (green, along k$_x$) and perpendicular (light blue, along k$_y$) to one of the electron pockets, as indicated by the colour-coded dashed lines in Fig. 1(c). 
Both panels (f) and (g) of Fig. 1 clearly show the dispersion relation of the Au-induced electronic states marked using the yellow dotted lines to be parabolic-like, with a band-bottom of a little under 150 meV below E$_F$ for both of the two perpendicular k-directions.
In the case of an ideal 1D system one would --- obviously --- expect no dispersion at all in one of the k-directions.
The electron pocket of interest (e.g. equivalent to that highlighted in Fig. 1(d)) is the most elongated along the direction of the green dashed line in Fig. 1(c).
If the paradigm of a system with a sufficiently 1D dispersion (so as to result in TLL behavior) is to be adhered to, one would consequently expect an insignificant dispersion along the k-direction traced out by the light-blue dashed line in panel (c), a conclusion that was indeed drawn in Ref. \onlinecite{meyer14prb}.
Fig. 1(f) shows that the high-resolution ARPES data from the high-quality Au-induced nanowire samples investigated here clearly contradict this expectation. 

This begs the question as to the relative orientation of the nanowires with respect to the dispersion relations shown in Figs. 1(f) and (g). Understandably, given the great promise of the Au/Ge(001) system as a 1D, TLL-system, this is a highly controversial point in the literature \cite{nakatsuji12natphy,blumenstein12natphy}.
However, seeing as this is a crucial point in the discussion of the dimensionality of the Au-induced states, we need to address the issue of the orientation of the nanowires with respect to the pair of orthogonal k-space cuts presented in Figs.\ref{fig:Fig1}(f) and (g) as carefully as possible.

\begin{figure*}
	\centering 
	\includegraphics[width = 18 cm]{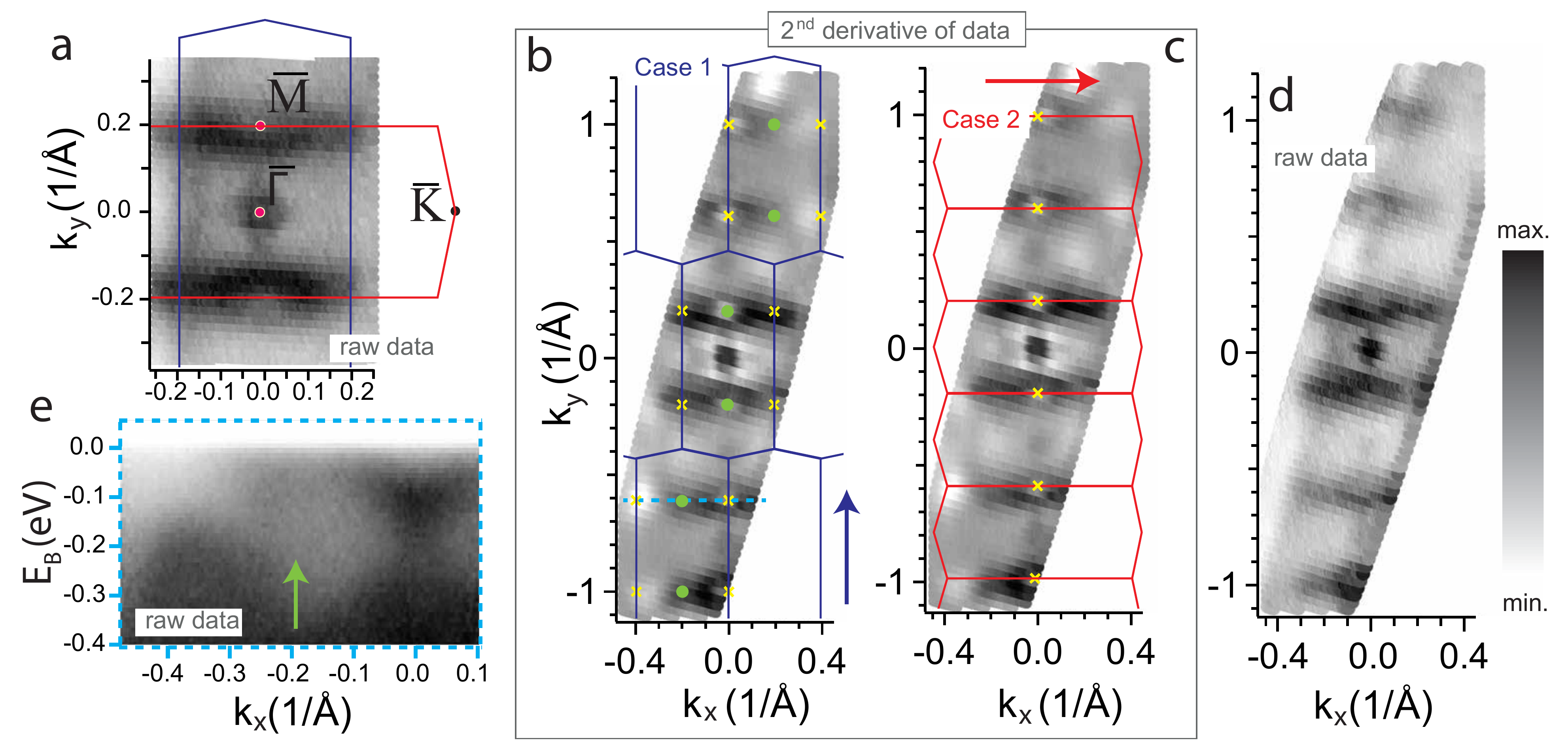}
	\caption{{\bf(a) ARPES  I(k$_x$,k$_y$) map with E$_B$=15 meV, recorded in the central portion the first surface Brillouin zone (SBZ). True-to-scale overlays of the expected SBZs from the two orthogonal nanowire domains such as those seen in STM (e.g. Fig. 1(a)) are shown using blue and red solid lines. It is clear that in the experimental geometry used, the ARPES intensity is dominated by one of the two nanowire domains. Panels (b) and (c) show a wide-k-range, constant energy map recorded at a sample temperature of 16K for a binding energy of 30 meV. As the contrast in the experimental data falls off (but remains non-zero) in higher Brillouin zones, the intensity scale shows the second differential (with respect to the binding energy), with the raw data being shown in panel (d). Superimposed on the two sets of identical experimental data in (b) and (c) are SBZs for the two possible nanowire orientations shown in a repeated zone scheme, with the green dots and x\textquotesingle s indicating equivalent points in the SBZ.  The blue/red arrows indicate the nanowire direction in each case. Panel (e) shows an I(k$_x$,E) cut through the raw data at the location in panel (b) indicated by the blue dashed line. All ARPES data were taken with a photon energy of 21.2 eV at a temperature of 20 K.
	}}
		\label{fig:Fig2}
	\end{figure*}

Fig. 2(a) shows another constant energy ARPES map for low-lying electronic states (here 15 meV below the Fermi level), measured from the same, high-quality Au/Ge(100) nanowire sample as from the data shown in Fig. 1. Despite the same dual-domain nature of the nanowire sample, the different geometry adopted in this experiment (compared to Fig. 1) results in a strong asymmetry in the intensity distribution between the two pairs of electron pockets - i.e. this experiment is more sensitive to the domains with nanowires running in one of the two possible orthogonal directions\footnote{In the experimental geometry relevant for Fig. 2, the $\Gamma$-$\bar{K}$ direction of one of the sets of surface nanowire domains is parallel to the entrance slit of the electron analyzer and antiparallel to the majority polarization vector of the partially linearly polarised VUV radiation. This results in favorable photoemission matrix elements for the nanowire states from one domain orientation, while the states from the orthogonal nanowire domains are evidently significantly suppressed.}. 
This intensity asymmetry is fortuitous, and makes these data a good representation of the electronic structure of a single nanowire surface domain. We do not know --- {\it a priori} --- from which of the two domain orientations these states originate, since both the ARPES light-spot and the LEED electron beam have lateral dimensions such that they necessarily cover multiple nanowire domains. Nevertheless, given the LEED pattern recorded from the same sample in the same ultrahigh vacuum system prior to the ARPES measurements shown in Figs. 1 and 2, we need only to distinguish between two different possible cases. 

\textit{Case 1:} With reference to panels (a) of Figs. 1 and 2, the blue SBZ is the correct one for states seen in the ARPES data. In this case, the band bottom of the electron pockets is not located at a high symmetry k-point such as the centre of an edge or a corner of the SBZ. This would be unusual, but not impossible. In this scenario, the Au-induced nanowire states would have the greatest velocity (i.e. smallest k$_F$, given a shared single band-bottom for both k-directions) for k along the nanowires. In the perpendicular k-direction (k$_x$ in Fig. 2(a)), the high-intensity regions of the electron pockets spill over into neighboring SBZ's, and if the 2D k$_x$-k$_y$ contours were not to close in this direction, it could be argued that they do form a quasi-1D state.

\textit{Case 2:} In this case, the red SBZ in Figs. 1a and 2a are the correct ones. This means that the band-bottom of the electron pocket is centred on the $\bar{M}$-point of the SBZ, indicated with a red dot in Fig. 2a. In this scenario, the Au-induced bands would not only form closed contours, but would also have their greatest velocity {\it perpendicular} to the nanowire direction. This situation would lead to the inescapable conclusion that the Au-induced states on Ge(100) are two-dimensional in nature. 

In the ARPES data from Au-induced nanowires on Ge(001) published to date, no clear intensity has been observed in the second BZ up till now \cite{meyer11prb,nakatsuji09prb,nakatsuji11prb,meyer14prb}.
In Fig. 2(d) we show the raw data of an I(k$_x$,k$_y$) ARPES map recorded over a wide k-range, with dark greyscale indicating high intensity just below E$_F$.
The clear, periodic repeat of the electron pockets yields, in total, six horizontal streaks of intensity, spanning beyond the first SBZ. 
To counteract the fall-off in contrast in higher SBZs, in panels (b) and (c) we plot the second derivative (with respect to the binding energy axis). Comparison of these two identical data images, each overlayed with the SBZ schemes pertinent to case 1 and case 2 yields a simple and effective manner to decide which case is correct. For the correct SBZ case, the k-space periodicity of the electron-pocket states and the SBZ scheme will match, with the electron pocket re-appearing at equivalent k-positions in higher SBZs.

In Fig. 2(b), the SBZ scheme that is overlaid on the experimental data belongs to case 1, for which the electron-pocket states could be argued to be quasi-1D (i.e. with greater band velocity parallel than perpendicular to the nanowires).
However, assuming this nanowire orientation, would mean that the band bottoms of the electron pockets would re-appear at completely inequivalent k-positions in the second SBZ, compared to the first SBZ.
In particular, for k$_x$=0, electron pocket band bottom appears at multiple k$_y$ values crossing the locations of both green dots and the yellow x\textquotesingle s.
This is not impossible, but would necessitate the presence of identical band bottom features centred at all the equivalent green dots and yellow x\textquotesingle s in Fig. 2(b).
To examine the credibility of this scenario, we show in Fig. 2(e) an I(k$_x$,E) image, taken along the dashed blue line in Fig. 2(b).
This line is chosen such that it cuts both a green dot, located at k$_x$= 0.195 \AA$^{-1}$, as well as the yellow cross at k$_x$=0, and at both these locations an identical parabola-like band should be seen, with a band bottom at ca. 150 meV E$_B$, like that seen in Fig. 1(f)).
Such a band-bottom is clearly visible in Fig. \ref{fig:Fig2}(e) for k$_x$ = 0 \AA$^{-1}$ but not at k$_x$= 0.195 \AA$^{-1}$ (green arrow). This inequivalence of the band structure at these two k-space locations excludes case 1 from being correct.
There is an extra, very faint, band bottom visible around 100 meV at k$_y$ = 0.23 \AA$^{-1}$ in Fig. 2(e). This state can also be observed in the constant energy maps of Fig. 1(c-e), located between 0.2 and 0.4 \AA$^{-1}$ on the vertical axis and is an extension of the electron pockets. From the data in Fig. 2e, it is not possible to determine whether there is a gap in k-space between this extra state and the main electron pocket centred in Fig. \ref{fig:Fig2}a and \ref{fig:Fig2}e at k$_x$ = 0 \AA$^{-1}$. This extra state will be discussed further in the next subsection, in which a tight binding model for the data is presented.

In Fig. 2(c), the red, repeated zone scheme SBZs superimposed on the ARPES constant energy map are linked to case 2, in which the Au-induced bands in fact show a higher velocity in the direction perpendicular to the nanowires than they do in the k-direction parallel to the nanowire-contrast seen in STM studies.
In this case 2, Fig. 2(c) shows the main electron pockets to be centred in a simple and periodic manner on the symmetry-equivalent $\bar{M}$ points (marked with yellow x\textquotesingle s) in all measured surface Brillouin zones.

Thus, from consideration of the matching of the fundamental symmetries of the surface crystal structure to the observed periodicity of the main features of the low-lying Au-induced electronic states resolved in low temperature ARPES experiments, the only possible conclusion is that the second case is the right one, and that the nanowires giving rise to the ARPES intensity in Fig. \ref{fig:Fig2} run along the direction shown by the red arrow in Fig. 2(c).
This means that the data shown in Fig. 2, together with those in Figs. 1(f) and 1(g) present a clear message. Rather than a quasi-1D E(k$_y$) landscape with a degree of coupling in the second, k$_x$ direction, the true situation for high quality Au-induced nanowires on Ge(100) is that of a 2D E(k$_x$,k$_y$) landscape in which the energy of the band forming the electron pocket changes faster with k$_y$, the k-direction \emph{perpendicular} to the nanowires than it does along the nanowires themselves.
From the high resolution ARPES point of view, the data presented here irrevocably exclude the Au/Ge(001) system from hosting a Tomonaga-Luttinger liquid. 
We note that a new synchrotron-based ARPES study of Au/Ge(100) recently also came to the same conclusion, namely that the electron pocket Fermi surfaces are closed, and thus two dimensional \cite{yaji16arxiv}  
\\
\subsection{Minimal tight-binding model}

To put the band structure observed in ARPES on a more quantitative footing, we formulate a minimal\footnote{Here, minimal means with the least number of non-zero hopping parameters.} single-band tight-binding model based on the c(8$\times$2) surface reconstruction. We start with the idealized lattice that corresponds to the	c(8$\times$2) structure seen in LEED and STM topography. Physically, the shortest-range hoppings should dominate and this is indeed what is found when simulating the data.
       
        \begin{figure*}
        	\centering 
        	\includegraphics[width = 18 cm]{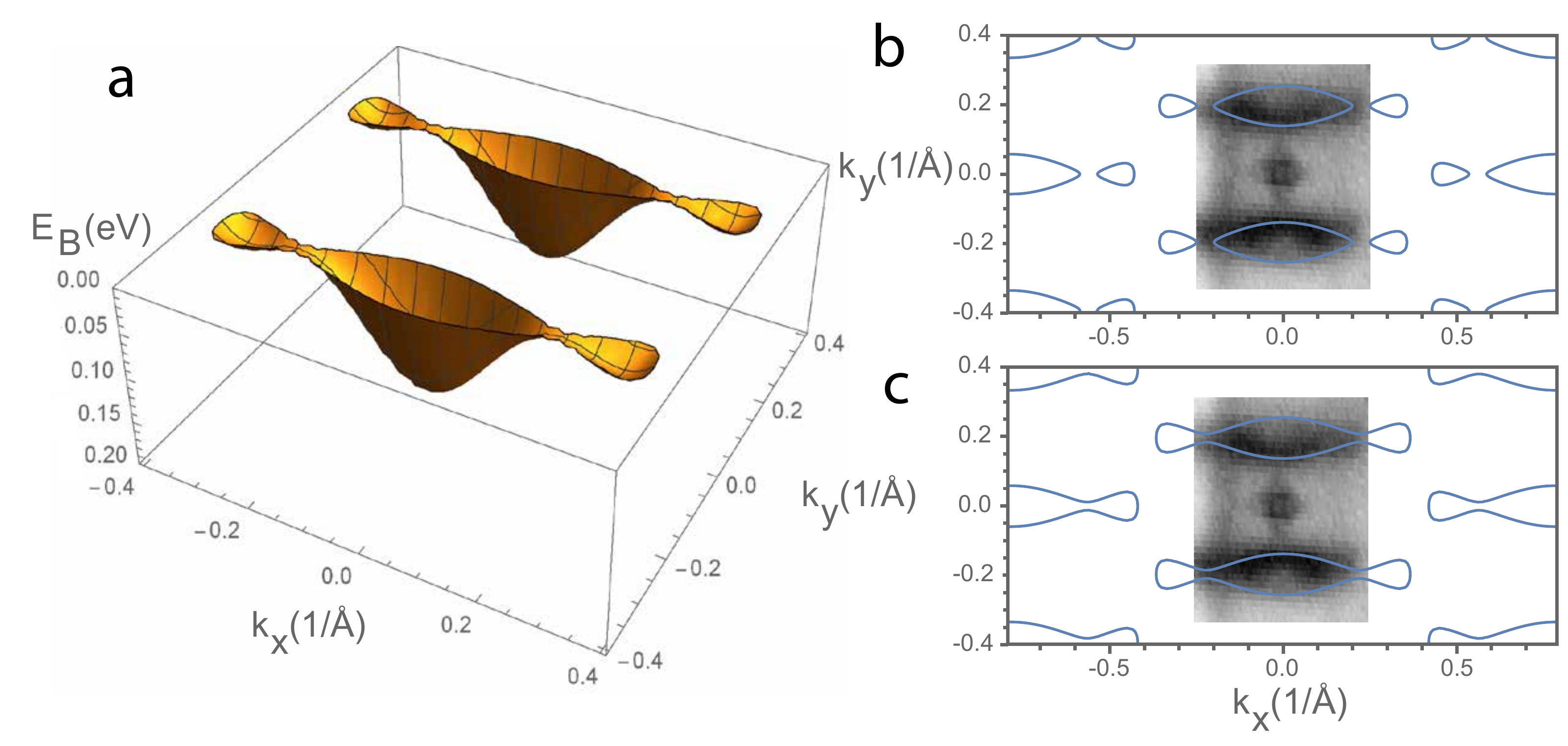}
        	\caption{{\bf(a) E(k$_x$,k$_y$) rendering of the
                    results of the tight-binding model (filled states
                    only) based on Eq. \ref{eq:3} with $\mu = -134$
                    meV and hopping parameters given in
                    Eq. \ref{eq:hoppings}. Panels (b) and (c) compare
                    constant energy surfaces in k$_x$,k$_y$ taken from
                    the tight binding model at the energies of 0 and
                    -10 meV in (a), with the ARPES data (E$_B$=15meV)
                    taken from Fig. 2a.}}
        	\label{fig:TB}
        \end{figure*}

We model the nanowire system as a two-dimensional lattice generated by the
Bravais lattice vectors $\vec{R}_{\pm} = (4\vec{e}_x \pm 16
\vec{e}_y)$\AA. We will use a pair of integers $(n,j)$ to label
site $j$ on nanowire $n$ located at $\vec{R} = n \vec{R}_+ +
j[\vec{R}_- + \vec{R}_+]$. The operators
$c^{\dag}_{\sigma,nj}$ ($ c_{\sigma,nj}$)
create  (annihilate) a fermion of spin $\sigma =
\uparrow,\downarrow$ on site $(n,j)$ and satisfy the canonical
anti-commutation relations $\{c_{\sigma,nj},c^{\dag}_{\sigma',n'j'}\}
= \delta_{\sigma\sigma'}\delta_{nn'}\delta_{jj'}$.
After exploring the space of spin-symmetric tight-binding models with
up to next-nearest neighbor hoppings that respect the symmetries of
the lattice we arrive at the model Hamiltonian
\begin{multline}
\label{eq:TB}
H = \frac{1}{2}\sum_{n,j} t_1^{\perp}(c^{\dag}_{nj}c_{n+1\, j} + c^{\dag}_{nj}c_{n+1\, j-1})  + t_1^{\parallel}c^{\dag}_{nj}c_{nj+1} \\ + t_2^{\parallel}c^{\dag}_{nj}c_{nj+2} +h.c .\, . 
\end{multline}
Here $h.c.$ denotes the hermitian conjugate. The resulting dispersion relation that we want to
compare to the ARPES data is then
\begin{multline}
\label{eq:3}
\varepsilon(k_x,k_y) = - \mu +2 t_1^{\perp}\cos( 4k_x)\cos(16 k_y) + t_1^{\parallel}\cos(8 k_x)\\ + t_2^{\parallel}\cos(16 k_x) .
\end{multline}
To obtain qualitative agreement with the experimental data, we have to take $t_1^{\perp}$
as the dominant hopping parameter, which a priori
seems surprising as it represents hopping between adjacent chains,
but it is a logical consequence of the higher velocity of the
Au-induced state perpendicular to the nanowires as described
in the previous section. Taking account of the band-bottom
energy and the qualitative features of the ARPES data we
arrive at a best estimate for the parameters
\begin{equation}
\label{eq:hoppings}
t_1^{\perp} \approx 130\ {\rm meV}, \quad t_1^{\parallel} \approx  65 \ {\rm meV}, \quad t_2^{\parallel} \approx - 45\ {\rm meV}
\end{equation}
hence with $t^{\perp}_1 \sim 2t^{\parallel}_1$. The chemical
potential in Eq. (\ref{eq:TB}) has to be chosen around
$\mu \approx -134 \ {\rm meV}$.

In Fig. \ref{fig:TB}(a), a 3D (E$_B$ k$_x$, k$_y$) representation of the calculated dispersion is shown. In panels (b) and (c) of Fig. \ref{fig:TB}, we compare tight binding (k$_x$,k$_y$)-contours shown as blue lines with ARPES data taken from Fig. 2(a). In the calculations E$_B$=0[10meV] in panel (b)[(c)].
The TB simulations show additional band minima - shallower pockets - on the sides of the main electron pockets. The intensity from these states is rather weak in the experimental data (e.g. see Fig. \ref{fig:Fig2}(e), for k$_x$ $\sim$-0.25). This makes a definitive call as to whether the experimental situation for E=E$_F$ is like the TB results shown in Fig. \ref{fig:TB}(b) or \ref{fig:TB}(c) difficult to make.

Naturally, the tight binding rendering of the E(k$_x$,k$_y$) landscape could be further fine-tuned by including longer range hoppings, however, further perfecting of the agreement with the experimental data 
at this level of approximation would be overkill, as more serious sources of potential discrepancy between the model and the ARPES data exist which are not
included in the tight-binding model such as the effects of electronic interactions, disorder and spin-orbit coupling.
We therefore leave more
refined studies of the band structure to more advanced methods
such as density
functional theory, although we note that the uncertainty as regards the exact structure of the Au-induced nanowire
system at the atomic level presently limits the applicability of even these more
sophisticated approaches.

For the purpose at hand, the
tight-binding model presented is sufficient to provide a
phenomenological parameter-set for the experimentally observed
$E(k_x,k_y)$ dispersion.
Two main points emerge from the analysis, namely that $t_1^{\perp} \approx 2 t_1^{\parallel}$, and that the TB contours are closed along the k$_x$-direction, not forming the modulated, yet continuous `tramlines' that would mark a quasi-1D fermiology of a candidate TLL system.

Thus, both the raw ARPES data itself, as well as a simple yet relevant minimal model of the underlying electronic states fail to support a 1D scenario for the Au-induced nanowires on Ge(100). In terms of the E(k$_x$,k$_y$) eigenvalues, Occam's razor points to the two dimensional character of the Au-induced nanowire states on Ge(100), despite the fact that their topographic signature in STM images looks so one dimensional.\\

      \begin{figure*}
      	\centering 
      	\includegraphics[width = 18 cm]{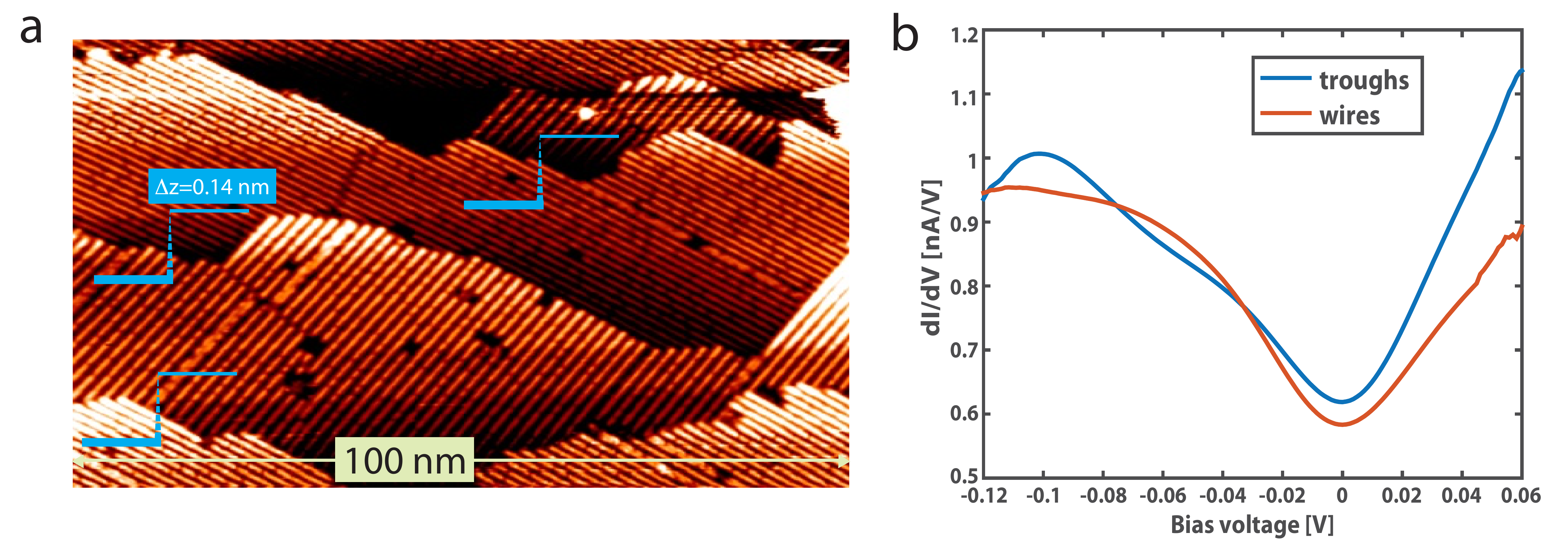}
      	\caption{{\bf(a) Large scale STM topographic image of the Au-induced nanowires on Ge(100) measured at room temperature with a bias voltage of -1.5 V and current set-point of 0.5 nA. Indicated in blue are the single-layer steps from one terrace to the next.   (b) $dI/dV$ spectra representative of the nanowires (red curve) and of the troughs between them (blue curve), extracted from a 20 nm $\times$ 20 nm area in the bulk of a nanowire patch (T = 4.7 K; setpoint current of 0.2 nA at a bias of -0.1V). 
      		}}
      		\label{fig:dIdVcurves}
      	\end{figure*}  

\subsection{STM/STS data}
In this section we turn to low temperature STS measurements carried out on identically-prepared Au-induced nanowire samples on Ge(001). As shown in Fig. \ref{fig:dIdVcurves}(a), the preparation protocol resulted in surfaces displaying large areas which are almost completely covered with Au-induced nanowires. The nanowires are perfectly straight and can have lengths of hundreds of nanometers, and at each single-layer step of the Ge(100) surface, the nanowire orientation rotates by 90 degrees.  For the STS measurements, nanowire regions with low defect densities were chosen from larger domains like those shown in Fig. 4a. Spectroscopy was recorded on a real-space grid consisting of either a 60$\times$60, 75$\times$75 or 100$\times$50 dI/dV pixels, with the pixel separation carefully chosen to yield an optimal compromise between measurement time and practically achievable spatial resolution. An I(V) spectrum was recorded at each pixel, while simultaneously recording the differential conductivity (dI/dV) with a lock-in amplifier.
In addition, from recent studies it is known that care must be taken when analysing STS data on the electronic structure of these nanowire arrays, since the troughs between the nanowires host electronic states not present on the nanowires themselves \cite{heimbuch12natphy,park14prb}. The data presented here were therefore analyzed carefully, so as to differentiate between spectra originating from atop or in-between the nanowire features seen in topographic data. The selection procedure that was used to separate spectra from the troughs and from the wires is described in detail in Appendix\ref{sec:slecetionSTS}.
	
Differential conductivities --- or local densities of states --- representative for on-nanowire measurements (red) and for measurements on the troughs between the nanowires (blue), are shown in Fig. \ref{fig:dIdVcurves}b.
The blue curve corresponding to the analogous data for the troughs displays a broad peak around -100 mV, which has been observed before \cite{heimbuch12natphy}, and attributed to a metallic state in the troughs.
Here we do note that this -0.1 eV LDOS peak does not show up in all differential conductivity traces recorded for the troughs.
Given the high degree of corrugation of these systems at the nm scale, an asymmetric or blunt STM tip could lead to averaging of the spectroscopic data between the troughs and wires, thus smearing out the trough-related LDOS peak at -0.1 eV. 

Besides this feature, both in on-wire and in-trough LDOS curves show a strong asymmetry between negative and positive bias.
This asymmetry is robust with respect to the junction resistance and is also independent of the details of any fine structure in individual LDOS curves.
In our STS measurements, this asymmetry is always clearly visible, independent of the temperature, location on the surface, and independent of the tip condition.
Consequently, the STS data shown in Fig. 4(b) are fully representative for tens of regions measured and thousands of individual dI/dV traces. 
This asymmetry is relevant, because as was explained in the introduction, TLL behaviour leads to an electron-hole symmetric suppression of the DOS around the Fermi level. Thus, our STS data are incompatible with the E/T scaling behavior reported in Ref. \cite{blumenstein12natphy}.

\begin{figure*}
	\centering 
	\includegraphics[width = 18 cm]{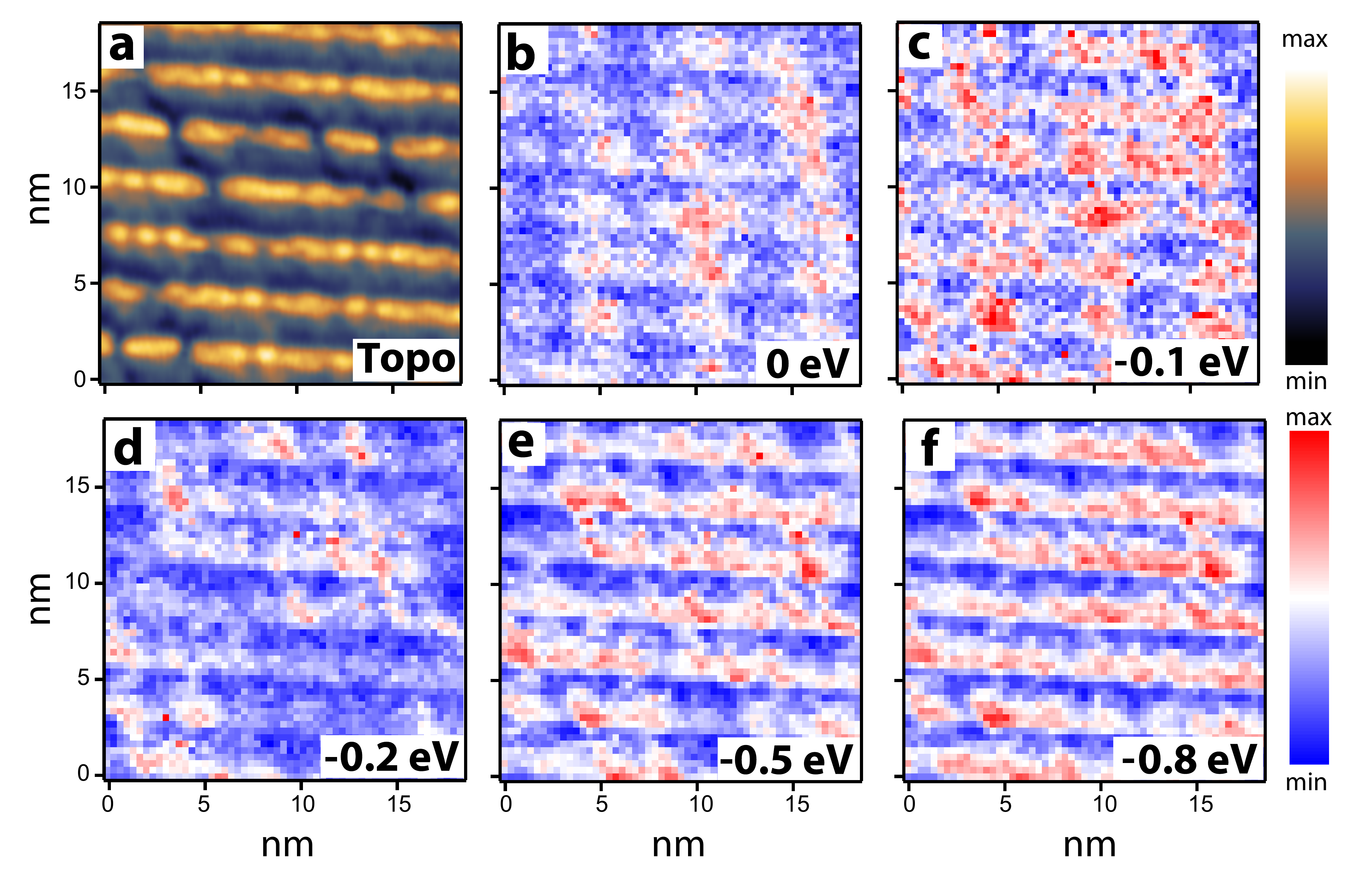}
	\caption{{\bf(a) Topographic image (bias 1 V, current 0.2 nA) of nanowire sample area for which the LDOS maps shown in (b-f) for different bias voltages were measured (set-point 1V and 0.2 nA). On the right the uppermost scale-bar belongs to the topographic image in panel (a), and lower one belongs to the LDOS maps shown in (b-f). All data were recorded at a temperature of 4.7 K.
	}}
		\label{fig:LDOSmaps}
		\end{figure*}

In Fig. \ref{fig:LDOSmaps} we show the spatial dependence of the LDOS for a straight and macroscopically defect/chain-end-free $\sim$20$\times$20 nm$^2$ section of Au-induced nanowires.
As reported in the literature \cite{heimbuch12natphy,park14prb}, straight conduction channels can be observed along the nanowire direction in the troughs.
These channels are most pronounced deeper in the occupied states, such as in panels e[f] for E=-0.5V[-0.8]V.
For bias closer to the Fermi level, the existence of straight conduction channels becomes less and less clear. Indeed, for E=E$_F$ (Fig. \ref{fig:LDOSmaps}(b)) and 0.1 eV below E$_F$ (Fig. \ref{fig:LDOSmaps}(c)) the spatial distribution of the LDOS is patchy, and could be interpreted either as a pearl chain of higher intensity patches running in a vertical direction, almost perpendicular to the nanowires\footnote{The slight angle of the pearl-chain-like structures with respect to the direction normal to the nanowires is comparable to the angle the VW reconstruction has in this direction \cite{park14prb}, suggesting a possible relation between this reconstruction and the enhanced LDOS patches. In addition, we note that the so-called bridge atoms, observed in the troughs in topographic images (see Fig. 3b of Ref. \onlinecite{safaei13prb} for an example in Twente-grown samples), are also possible candidates for the origin of these patches of higher LDOS. When summed over space, these LDOS patches yield the  LDOS peak observed at a bias of -0.1 eV in the trough-averaged STS spectra shown in Fig. \ref{fig:dIdVcurves}(b).}, or as LDOS patches along the troughs between nanowires. This either/or situation essentially expresses the 2D nature of these low lying occupied states.
Comparing to our ARPES data, we remark that the highest intensity in ARPES for the Au-induced electron pockets is close to 100 meV binding energy, and the fact that these electronic states disperse more rapidly perpendicular to the nanowires would seem to mesh with the fact that these patches of high intensity in the low energy LDOS maps also run in the direction perpendicular to the nanowires.
Similar results have been found by other groups \cite{park14prb}, whereby straight differential conductance channels running along the nanowire direction were only observed for negative biases V$_{b}\le$-200 mV.

The STS data presented here and in Ref.  \onlinecite{park14prb} lay bare a discrepancy with those of Ref. \onlinecite{blumenstein13jpcm} in which the straight features in the conductance were linked to the electron pockets seen in ARPES.
In past discussions as the the dimensionality of the nanowire-related electronic states, differences observed between the energies of the possible 1D states in ARPES and STM data have been suggested to be caused by a shift in the chemical potential, for instance induced by the differences in doping levels of the different substrates used \cite{schafer09prl}.
However, here we present data taken with ARPES and STM/STS on identically-prepared samples, both using the same batch of substrates from a single Ge(100) wafer.
The ARPES data (e.g. Fig. \ref{fig:Fig1}) enables to easily determine the position of the chemical potential, and no significant shifts are observable between the ARPES data presented here and analogous data reported using differently doped substrates in the literature \cite{nakatsuji09prb,meyer11prb,nakatsuji11prb,meyer14prb}.
Therefore differences in chemical potential or even variation in the details of the sample preparation protocols cannot be used to argue that the electronic states seen within 100meV of the Fermi level in ARPES are 1D, yet the differential conductance images in STS display essentially 2D patterns when imaged in the same low energy region.

From the above, it is clear that the 1D-like conductance channels observed by STS in the nanowire troughs at higher bias energies (e.g. Fig. \ref{fig:LDOSmaps}(e,f)) do not come from the low energy electron pockets observed with ARPES. The question is then where do these straight conduction channels come from? Here we propose two alternative explanations:

1)  The ARPES data presented here show the presence of both Au-induced surface states, and 2D states derived from the bulk germanium bands.
In Ref. \onlinecite{Ohtsubu13prb} it was shown that for interfaces between various metals and Ge(111), surface states and resonances are created with maxima in their charge distributions lying between 5 and 10 layers below the surface.
If the situation for Ge(100) were to be similar, sub-surface, 2D Ge states could exist, and, as the tops of the nanowires are at least several germanium layers higher up than troughs between the nanowires \cite{houselt08prb,kockmann09jpcc}, the STS signal measured in the troughs would be much more likely to pick up a contribution from these germanium subsurface states, thus yielding higher conductance stripes running along the troughs. Put simply: such quasi-2D, Ge subsurface states could `shine through' in the troughs between the nanowires. 
In appendix \ref{sec:kzdep}, photon energy dependent ARPES data from Au/Ge(100) shows that the electronic states up to as far as 1.8 eV below E$_F$ are essentially independent of k$_z$, and thus are of 2D character, which would be consistent with the scenario sketched above. 

2)  Another possible explanation is also closely related to the high degree of corrugation such Au/Ge(100) nanowire surfaces display.
A system with spatially inhomogeneous LDOS, but also at the same time strongly varying height profile such as is the case here makes it all but impossible to decouple height and LDOS information in the STS signal.
All published work on this system agrees that these samples possess features with a height difference of several atomic layers spatially separated by only 1.6 nm.
This is an extremely challenging situation for STM/STS mapping using a real-life tip, the extremity of which may be smaller than, of the same order, or larger than the inter-nanowire trough size.
This makes in particular the spatial dependence of STS measurements\footnote{Please recall that the electron-hole asymmetry of the atop and in-trough DOS curves from STS \emph{was} independent of tip conditions/sharpness.} highly dependent on the sharpness and shape of the tip.
Consider setting up an LDOS map with the tip set atop a nanowire --- here the choice of set-up bias voltage and current will really set the height difference between the tip and the sample. 
However, when the tip is set at a trough, the shortest distance from the sample to the tip for a given set-point is less well defined --- particularly if the tip apex and the trough profile were to match (like two gears). In such a case, it is easy to see how lateral tunnelling could also take place, boosting the final dI/dV signal for reasons other than an enhanced LDOS at the bottom of the trough itself.

This discussion serves to show that a degree of caution is required in the interpretation of STS data from these systems, a caveat that does not apply to the `remote probe' of the occupied states provided by ARPES experiments.\\

\section{Discussion}

In this section we discuss our findings in light of earlier research performed on Au/Ge(001). First we note that the basic structure of the Au/Ge(001) nanowires observed with STM is always the same, even though the samples are prepared in different ways with differently doped germanium substrates \cite{wang04prb,schafer08prl,houselt08prb,kockmann09jpcc,blumenstein11prl,niikura11prb,blumenstein13jpcm,park14prb}.
Additionally, the observed LEED patterns  are  the same in every study that published LEED data on the Au/Ge(100) system \cite{schafer08prl,schafer09njp,nakatsuji09prb,blumenstein11prl,meyer11prb,nakatsuji11prb,meyer12prb,blumenstein13jpcm,lichtenstein15ss}, always showing a c(8$\times$2) reconstruction with a (8$\times$4) superstructure.
While the exact nature of this structure on the atomic level is still under debate, it is clear that all groups are studying the same Au-induced surface reconstruction.

In contrast to the observations concerning the structure of the Au/Ge(100) system, the dI/dV spectra measured with STS show strong variations between groups.
While the data presented in this manuscript and previous studies all agree that there is a suppression of the density of states at the Fermi level, the exact shape of this dip is by some measured to be particle-hole symmetric around zero bias voltage \cite{blumenstein11natphy} while in other data, such as that presented here and in Ref. \onlinecite{park14prb}, an symmetric dip is observed - i.e. there is no sign of p-h symmetry. Furthermore, the supposedly 1D conductance channels observed by STS mapping are seen in some datasets \cite{blumenstein13jpcm} to be at low bias voltages (ca. $\pm$ 20 meV from the Fermi level), in others at negative biases of 100 meV \cite{heimbuch12natphy}, and in the data presented here and in Ref.  \onlinecite{park14prb} only for negative biases exceeding 200 mV.
Given the consistency of the LEED and topographic data across samples and groups, and the fact that there is agreement on the band bottom values for the Au-induced electron pockets ARPES across samples and groups \cite{schafer08prl,nakatsuji09prb,meyer11prb,nakatsuji11prb,blumenstein13jpcm,meyer14prb}, differences in the position of the chemical potential in different and differently prepares samples can be ruled out. 
This means that the strong variation of the spatial distribution of low energy LDOS from one study to the next   is most likely to be related to the strong influence of the tip-shape on the STS measurements due to the large height corrugation resulting from the closely spaced nanowires, and due to the possibility of tunnelling from sub-surface, 2D Ge-related states when the STM tip is in the troughs between the nanowires.      

We now discuss our ARPES data, compared to those published by different groups.
As mentioned above, similar bands structures are observed, yet there are large differences in interpretation of the data.
In all ARPES data, Au-induced electron pockets are observed with their band bottoms located close to (k$_x$,k$_y$) = (0.2,0), (0.2,0), (0,-0.2) and (-0.2,0) \AA$^{-1}$, with a band bottom energy of little below 150 meV binding energy \cite{schafer08prl,nakatsuji09prb,meyer11prb,nakatsuji11prb,blumenstein13jpcm,meyer14prb}.
The disagreement in interpretation boils down to two important issues:
1) The exact shape of the electron pockets at the Fermi level together with the hopping parameters in two orthogonal k-directions and
2) The direction of the nanowires observed in STM topographic studies, compared to the k-space coordinates on which the ARPES data are presented.
In Refs. \cite{schafer08prl,meyer11prb,blumenstein13jpcm}, the electron pockets are interpreted as having a straight, half-pipe-like dispersion relation, with no states `closing' this quasi-1D Fermi surface in the k-direction perpendicular to the nanowires. In a later publication \cite{meyer14prb}, this picture was refined and mention is made of a slight dependence of the nanowire states on t$_{\perp}$, but this was argued to be insufficient to endanger TLL behavior. 
In Refs. \onlinecite{meyer11prb,blumenstein13jpcm,meyer14prb}, the electron pockets observed with ARPES are associated to the straight features observed in STS conductance maps --- a point to which we will come back in the next paragraph.
 
Our ARPES data presented here agree well with the interpretation put forward in Ref. \cite{nakatsuji11prb} in which data taken on samples with a slight miscut to the (001) plane seemed to show a surface state which was 2D in nature, a conclusion reinforced by recent ARPES data from the same group \cite{yaji16arxiv} .
In our experiments, we clearly resolve the two-dimensional dispersion relation of these electron pockets, with tight-binding simulations showing a difference in the magnitude of the orthogonal hopping integrals of a factor $\sim$2, which is not even close to a quasi-1D limit.
Additionally, as our data allowed the clear identification of the electron pockets in the second and third surface Brillouin zone, the direction of the nanowires as seen in STM could be unequivocally linked to the k-space co-ordinates of the ARPES data.  This leads to only one possible conclusion that the ARPES dispersion for the k-direction perpendicular to the nanowire-contrast seen in STM is the strongest. The ARPES-measured dispersion parallel to the nanowires is less, but is still significant - resulting in a closed, 2D Fermi surface contour. 

We conclude that both the ARPES and STM/STS data presented here are unable to support a one-dimensional scenario for the gold-induced metallic states in the Au/Ge(001) nanowire system, thus disqualifying this material system as a possible host for TLL physics.
  
 In view of future searches for TLL physics at surfaces of solid-state systems we remark here on the effects of finite
 chain lengths, higher dimensional coupling and disorder which limit the
 temperature and energy window in which TLL behavior may be
 observed. As experimental confirmation of TLL physics often focuses  on observations
 around the Fermi level, this presents a certain tension between
 theory and experiment. In particular, the low-energy response of quasi one-dimensional systems may be
 dominated by weak magnetic or charge density order in two or three
 dimensions, or may become Fermi-liquid-like. Neutron scattering data
 on materials hosting quasi-1D spin systems are illustrative in this respect.
 Inelastic neutron scattering experiments on the quantum spin-ladder
 material (C$_5$H$_{12}$N$_2$)CuBr$_4$ \cite{2009_Thielmann_PRL_102}
 and the spin-1/2 Heisenberg chain material CuSO$_4$5D$_2$O 
 \cite{2013_Mourigal_NATPHYS_9} are well-described in terms of
 one-dimensional models, which in the pure theory can be shown to be in a spin-Luttinger-liquid phase. While
 the agreement to the experimental data above a certain threshold energy is
 remarkable, the correspondence is lost below a threshold energy due to 3D ordering. Another enlightening example comes from the phase diagram of certain transition metal compounds or Bechgaard salts,
 which show a multitude of ordered phases at low temperatures and even
 Fermi-liquid-like behavior, while at temperatures above the critical
 temperature --- which can be as high as 100 K --- one finds a TLL
 phase \cite{1999_Bourbonnais_inbook}.  In fact, also Au/Ge(001) is known to exhibit a  high-temperature transition at 585 K.  Above  this temperature the nanowires display a higher degree of 1D structural order, characterized by simple dimer buckling, while below the STM images indicate a more glassy superstructure with the characteristic VW shapes and the appearance of complex inter-chain correlations \cite{blumenstein11prl}.
 The electronic conduction channels of a surface system like
 Au/Ge(001) will therefore be likely to suffer from similar instabilities towards 2D
 or 3D ordering or will show a dimensional crossover at some energy
 scale, even if they were to host weakly coupled 1D conduction channels.
 A rough estimate for such a scale may be determined from
 Renormalization Group arguments (see Appendix C). The typical lowest temperatures for STS on
 Au/Ge(001) are of order 5K (e.g. data presented here and in Ref. \onlinecite{blumenstein11natphy}). This temperature also roughly corresponds to the temperature
 at which quantization effects due to the finite chain lengths of
 about 100 nm could start to obscure TLL physics, if we were to assume the Au-induced electronic states of Au/Ge(001) to be 1D. Entertaining the
 one-dimensional interpretation with $K_c\sim 0.26$ as proposed by
 Ref. \onlinecite{blumenstein11natphy}, and estimating quantities
 such as the Fermi velocity from the ARPES data in their
 interpretation, we would arrive at an estimated admissible
 inter-chain hopping $t_{\perp}/t_{\parallel} \sim 0.1$ (see App. \ref{app:RG}), which is of course an order of magnitude away from the factor of two that we have obtained from our direct ARPES determination of the E(k$_x$,k$_y$) landscape.

 We now return to the remaining puzzle: the incoherent nature of the electronic
 states close to the Fermi level seen in ARPES (see App. \ref{app:dip}), and the observation
 of an anomalous suppression in the density of states at the Fermi level in the STS
 data (see Fig. \ref{fig:dIdVcurves}(b)).
 Both these puzzling observations are found in  \emph{all} published data on Au/Ge(001).
 On the one hand, the dip in the
 LDOS at E$_F$, and in particular the observation of its universal scaling with both
 temperature and energy in Ref. \cite{blumenstein11natphy} has formed the
 strongest argument in favour of TLL physics in Au/Ge(001)
 nanowires.
 On the other hand, the ARPES data presented here rule out
 TLL-physics for Au-induced nanowires on Ge(001), as
 their electronic states within 100 meV of the Fermi level are
 unequivocally shown to be 2D in character. This conclusion is also
 supported by the STS data from identically-fabricated
 samples. Therefore, the issue of what else could give rise to the marked
 departure of the spectral function and (local) density of states
 from the regular metallic paradigm of the Fermi liquid is one that warrants discussion, which we provide in the following from a
 theoretical point of view. 
 
 In Ref. \onlinecite{blumenstein11natphy} two other possible
 explanations are considered for the suppression of the density of states at the
 Fermi level besides TLL behaviour,
 namely a Coulomb pseudogap and dynamical Coulomb blockade. The
 Coulomb blockade is set aside since the experimentally obtained
 resistance of the tunnelling circuit does not meet the theoretical
 requirements. We agree that this mechanism for the zero-bias
 anomaly, which in its standard form relies on the impedance of the
 tunnelling circuit, would certainly not explain the corresponding ARPES
 data.
 As a general mechanism, however, we regard the interplay of
 disorder and interactions effecting the density of states at the
 Fermi level as the most likely cause for the observed DOS
 suppression in what our data show to be a 2D system, a conclusion that echoes that made in Ref. \cite{yaji16arxiv}.
 This is the
 physics behind the Coulomb pseudogap, which comes in two varieties
 depending on whether the system is insulating or remains
 metallic. In Ref.  \cite{blumenstein11natphy} the metallic
 Altshuler-Aronov anomaly \cite{1979_Altshuler_SSC_30} is discussed
 and dismissed because it predicts exponential behavior close to the
 Fermi level \cite{2002_Bartosch_EPJB_28,2001_Mishchenko_PRL_87}
 which did not fit the data of Ref.  \cite{blumenstein11natphy}.
 This is true, if one considers a system in one dimension, but for effectively two-dimensional systems one
 generically gets a linear dependence for low energies
 \cite{2002_Bartosch_EPJB_28,1998_Kopietz_PRL_81}. However,
 variations of the theory show that general power-law exponents are also
 possible \cite{1998_Kopietz_PRL_81,2001_Egger_PRL_87}. The linear
 suppression of the DOS is also generic on the insulating side of the
 metal-insulator transition, for which Efros and Shklovskii predicted
 a soft gap due to Coulomb interactions and disorder
 \cite{1975_Efros_JPC_8}.  As far as the tunnelling experiments are
 concerned, it is useful to consider parallel discussions on
 universal scaling in transport phenomena in organic conductors
 \cite{2010_Kronemeijer_PRL_105,2015_Abdalla_NATURE_5}.  In those
 systems, the conductance is considered to be caused by hopping
 between localized states --- which corresponds to the conduction
 mechanism for disordered materials in the Efros-Shklovskii class ---
 and it has been shown that that transport in such systems can give
 rise to universal scaling with temperature and bias, emulating the
 TLL predictions
 \cite{2010_Rodin_PRL_105,2014_Li_JAP_116,2013_Asadi_NATCOMM_4}.  It
 is an open question whether universal scaling for the density of
 states can be obtained from disorder-based theories.\\
 
\section{Conclusions}

We have performed an extensive experimental study of the self-organized Au/Ge(001)
system --- commonly being referred to as being composed of nanowires
--- using LEED, ARPES and STM. Our high resolution ARPES data clearly shows a dependence of the low-lying Au-induced electronic states on two orthogonal directions in momentum.
The observed k-space periodicity of the ARPES data fixes irrefutably the orientation of the surface Brillouin zone with respect to the nanowires, showing that the relevant bands have their highest velocity perpendicular to the nanowires.
Moreover, the observed periodicity in k-space cannot be matched with a quasi one-dimensional Fermi surface, and this additionally underpins the form of the ARPES I(E, k$_x$, k$_y$) images which show that this system supports two-dimensional states in which the low-lying electron
pockets form closed Fermi surfaces.

Considering a simple tight-binding
model based on the c(8$\times2$) reconstruction as the relevant
surface symmetry --- consistent with observed LEED pattern, STM data
and ARPES data --- we find that the qualitative features of the ARPES
data are reproduced quite naturally by short-range hoppings.
This
allowed the hopping to be quantified along and perpendicular to the nanowires, and we
found that the latter is the larger, by a factor of
two.

In keeping with this, the bias dependence of the spatial maps of the LDOS from STS experiments agrees with a lack of 1D character for the low-lying, Au-induced electron pockets observed in ARPES.
The STS spectra measured in the troughs show a broad peak around -0.1 V bias voltage, which we show to be likely to be associated with enhanced LDOS patches observed in the maps. These patches resemble pearl-chain-like structures oriented almost perpendicular to the nanowire direction and are most clearly resolved at the Fermi-level. 
These observations from STS agree well with the conclusions from ARPES of the dominance of electronic hopping perpendicular to the nanowire direction. 

Taken together, all these findings prohibit the observation of one-dimensional physics at low energies in
these materials, and thus also exclude the existence of a
Tomonaga-Luttinger liquid in the nanowire samples.
The density of states close to the Fermi level observed
in both tunneling data and the k-integrated photoemission data is anomalously
suppressed. As this cannot be connected to TLL physics, it is most likely an Altshuler-Aronov-like effect, caused by the interplay of disorder and interactions in a two-dimensional metal. Several theoretical
studies indicate that the apparent universal scaling of the tunnelling density of states with temperature and bias can in fact be due to such disorder+interactions+2D effects,
but in fairness it should be stated that a quantitative theoretical underpinning for the observed
suppression of the density of states remains elusive at this point.\\

\begin{acknowledgments}
	We would like to thank the Stichting voor Fundamenteel Onderzoek der Materie (FOM, 10ODE01, 10ODE02 and 10ODE03), the Nederlandse Organisatie voor Wetenschappelijk onderzoek (NWO/CW ECHO.08.F2.008) and the MESA+ Institute for Nanotechnology (HW) for financial support.\\
	We thank HZB for the allocation of synchrotron radiation beamtime for the photon energy dependent ARPES experiments the data from which are shown in App. \ref{sec:kzdep}, and acknowledge A. Varykhalov for support during the beamtime. 
\end{acknowledgments}

\newpage
\setcounter{section}{0}

\section*{Appendices}
\subsection{k$\bm{_{z}}$ dependence of the ARPES data of Au/Ge(100) }\label{sec:kzdep}
\setcounter{figure}{0}
\makeatletter 
\renewcommand{\thefigure}{A\@arabic\c@figure}
\makeatother

The ARPES measurements presented in this appendix were performed at the UE112-PGM-2a-1\string^2 end-station of the BESSY II synchrotron radiation facility at HZB using a Scienta R8000 hemispherical electron analyzer and a six-axis manipulator.\\

\begin{figure}[h]
	\centering
	\includegraphics[width=0.5\textwidth]{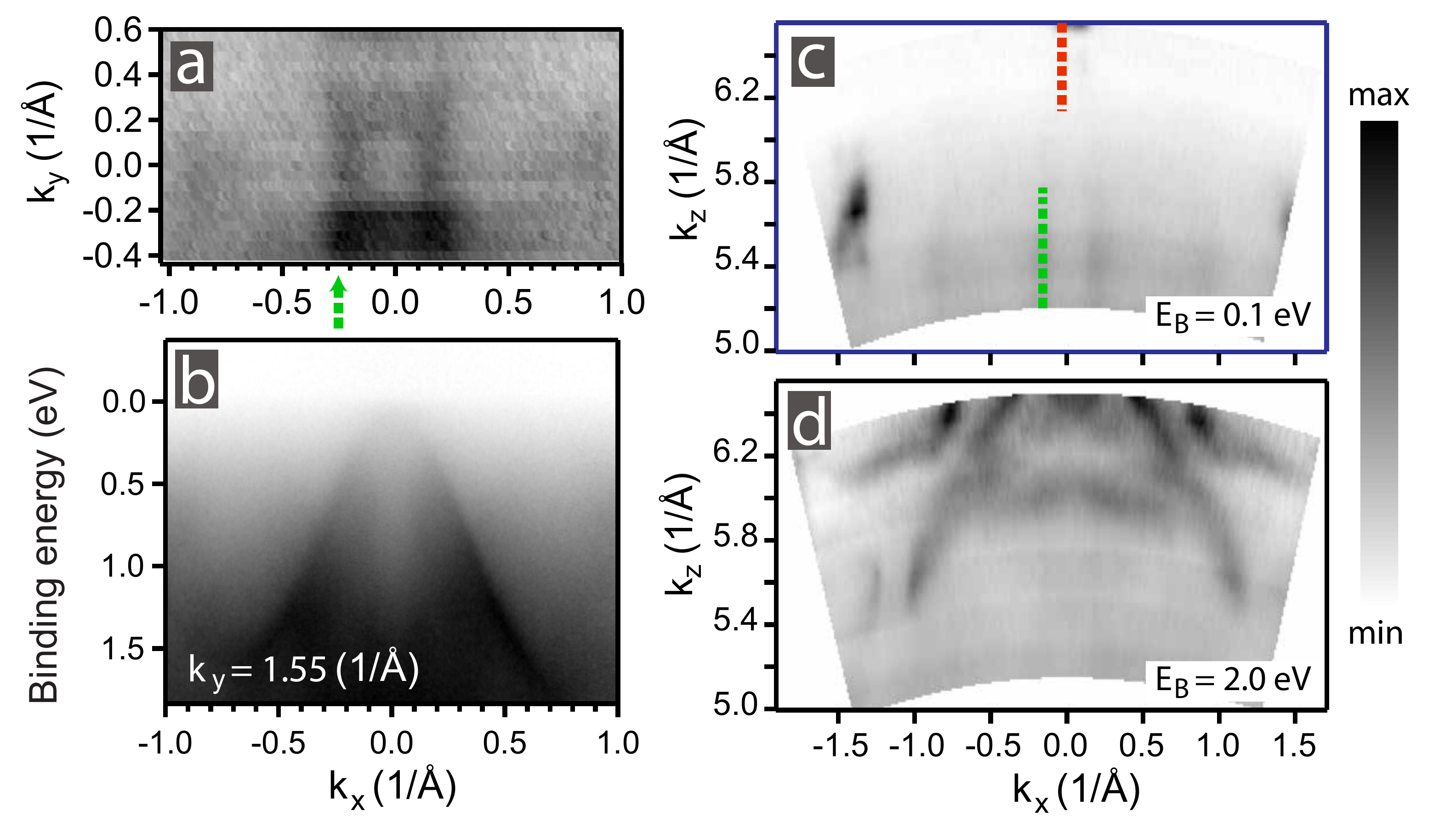}
	\caption{{\bf (a) I(k$_x$,k$_y$) map for E$_B$=10 meV. (b)
            I(E,k$\bm{_x}$) at k$\bm{_y = 1.55}$ (close to
            $\bm{\Gamma}$ in the second Brillouin zone in the extended
            zone scheme). Both panels (a) and (b) are recorded with
            $\bm{h\nu = 100}$ eV. (c) and (d) show
            $\bm{h\nu}$-dependent datasets, plotted as I(k$_x$,k$_z$)
            at constant binding energies of 0.1 and 2 eV binding
            energy, respectively. The red and green dashed lines are guides
            to the eye for the straight features observed close to the Fermi level which are related to bulk-derived and Au-induced states, respectively. All measurements were performed at 30K.}}
	\label{fig:kzdep}
\end{figure}

Figs. \ref{fig:kzdep}(a) and (b) show the Au-induced electron-pocket states and Ge bulk-derived bands respectively, measured with a photon energy of 100 eV. These measurements were carried out on samples that were slightly aged after the transportation from Amsterdam to Bessy II in Berlin in a UHV suitcase (p $< 5\times10^{-10}$). At the synchrotron, the samples were regenerated by direct current annealing at 650 ($\pm$ 25) K. The Au-induced, electron pocket surface states are not as clearly resolved in these data as in the measurements carried out just after sample preparation and under ideal vacuum conditions in the Amsterdam lab (shown in the main body of the paper). Nevertheless, the data in Fig. A1 are sufficient to enable tracking of the dispersion of the electronic states as function of photon energy, as shown in Figs. \ref{fig:kzdep}(c) and (d). Here we present I(k$_x$,k$_z$) at E$_B$ = 0.1 and 2.0 eV, respectively. At energies close to the Fermi level (panel [c]), we observe no k$_z$ dispersion for the Au-induced, electron pocket states (see green guide-line), and also no k$_z$ dispersion for the states close to the $\Gamma$-point, which are bulk-derived states of Ge character. In contrast, deeper in the valence bands, such as the 2 eV E$_B$ data shown in panel (d), bulk-like, 3D dispersion of the Ge-bands is clearly seen, indicating that our resolution in k$_z$ is sufficient. Earlier work on the Au/Ge(001) system \cite{meyer14prb} also showed the low-lying, Ge-derived bands to be two dimensional. The surface nature of these states has been argued to be related to the observations of sub-surface states at Ge(111)/metal interfaces \cite{Ohtsubu13prb}, which originate from the bulk valance band, perturbed by the interface formed with different metals. The authors of Ref. \onlinecite{meyer14prb} suggested that a similar mechanism is behind the creation of sub-surface states in the Au/Ge(001) system as well, which does explain why these states do no show any k$_z$ dependence but do follow the bulk valence band closely. This interpretation fits well with our data, we can add that for binding energies in excess of $\sim$ 1.8 eV, a cross-over point is reached where the states start to show a clear dispersion in k$_z$ (see Fig. \ref{fig:kzdep}), making an attribution to pure bulk germanium bands the simplest interpretation. \\

\subsection{Selection procedure STS curves}\label{sec:slecetionSTS}
\setcounter{figure}{0}
\makeatletter 
\renewcommand{\thefigure}{B\@arabic\c@figure}
\makeatother

\begin{figure}[h]
	\centering
	\includegraphics[width=0.5\textwidth]{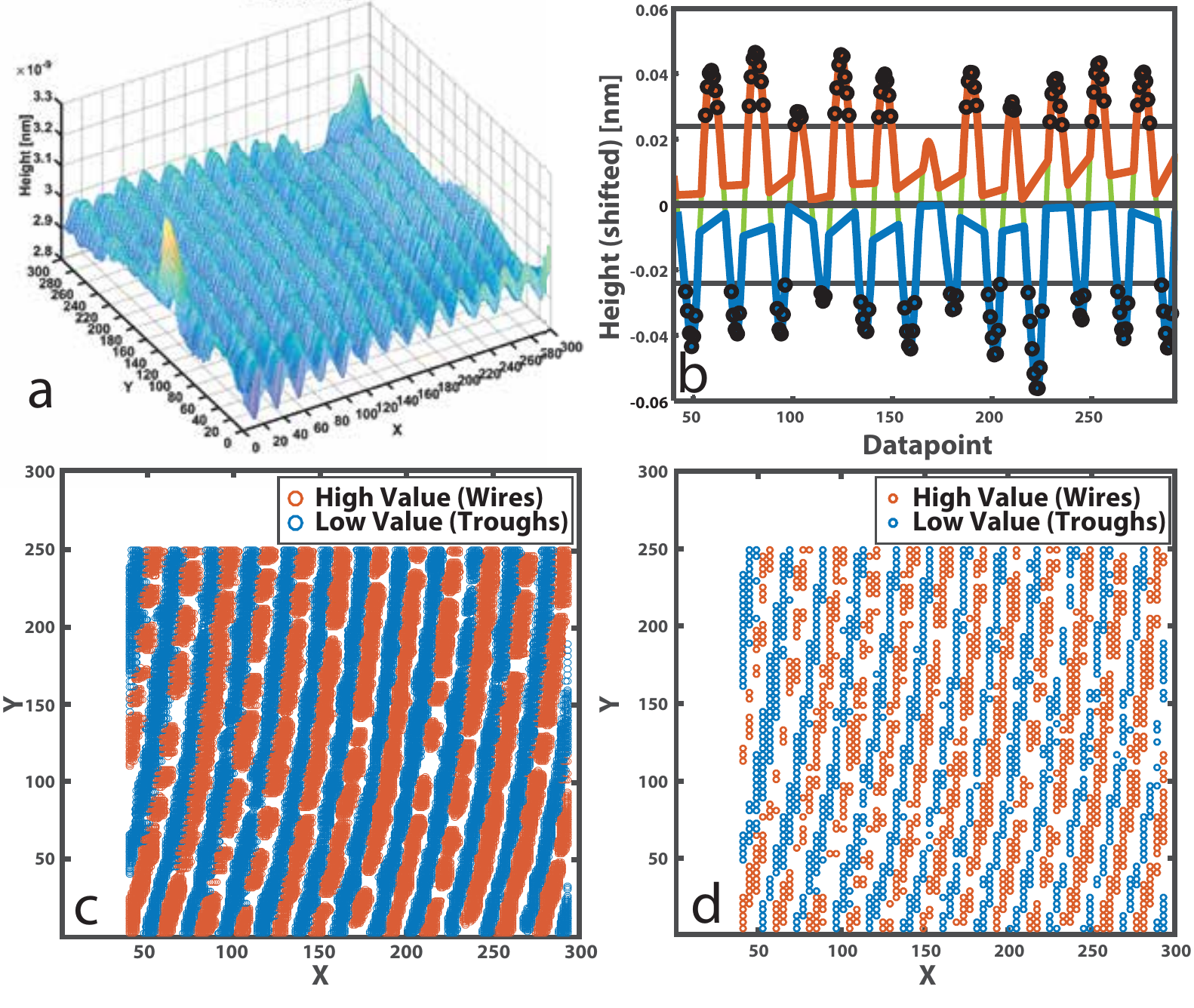}
	\caption{{\bf Illustration of the data selection routine for spectroscopic grid maps, recorded on the Au/Ge(001) system at low temperatures. (a) Topographic information of the scanned area. (b) Example of definition of data points as high or low from line scan height profiles. (c) Map of the high[low] values from the topograph. (d) Map of the spatial pixels matching locations at which dI/dV STS curves were recorded. Red[blue] pixels correspond to high[low] topographic height values. See text for a step-by-step description of the procedure.}}
	\label{fig:wirescript}
\end{figure}

The STM/STS data were analyzed by first importing the topography (see Figure \ref{fig:wirescript}(a)) and selecting a suitable region influenced as little as possible by defects on the surface.
The topography is then binned into high values and low values (corresponding to wires and troughs) by comparing the points (x) of each line (y) to specified threshold values (Figure \ref{fig:wirescript}(b)). For this, each line scan was individually levelled to its average value and the deviation of each point above or below this value was extracted. The high[low] thresholds are then defined as a specific percentage of the mean of all points above[below] the global average for the line scan. Only data points above[below] these thresholds are passed on. This procedure ensures that every line is treated individually and the selection is not influenced by global effects such as surface slope, drift or other height correlations. High[low] data (see Figure \ref{fig:wirescript}(c)) is then compared to the coarser spatial grid on which the STS spectroscopy map was recorded. Each STS trace whose spatial location matches a high[low] location, is then assigned to the high[low] group of spectra (see Figure \ref{fig:wirescript}(d)). The minima of all the differential conductivity curves are subsequently shifted to 0 V, to correct for a known\footnote{Comparison of dI/dV data from the lock-in amplifier with numerically calculated differential conductivity from the I(V) curves, proves this effect to be due to the lock-in amplifier.} artefact in the lock-in amplifier. The high[low] datasets are then averaged over all the matching data points of the selected region.\\

\subsection{Renormalization Group arguments }
\label{app:RG}
In this appendix - ignoring for the moment the experimental data we present in the main body of the paper to the contrary -  we entertain the one-dimensional scenario for
Au/Ge(001) and imagine that the Fermi surface is approximately
straight with the highest velocity of the electronic states parallel to the wires. Our aim here is to estimate the temperature and energy scales for the different
transitions expected for a TLL.  The temperature scales could be important for
systems such as Au/Ge(001) in relation to higher dimensional
coupling, and disorder can be obtained from RG based arguments
described in detail Ref. \onlinecite{GiamarchiBOOK} and references
therein. We outline this reasoning here applied to the Au/Ge(001) nanowires.

The starting point is an infinite array of one-dimensional wires each described as a TLL with charge and spin velocities $v_c$ and $v_s$ and Luttinger parameters $K_c$ and $K_s$. We assume spin-isotropic interactions, $K_s=1$, and adhere to the reported $K_c \approx 0.26$ \cite{blumenstein11natphy} . 

For finite chain length $L$, quantization effects may obscure the TLL behavior if the thermal length $L_T \sim v_c/T$ (in units such that $\hbar=k_B=1$) becomes comparable to $L$.
In Au/Ge(001) the maximal nanowire length is approximately $L\sim 100$ nm.
From the maximal dE/dk at the Fermi level in the ARPES data which is of
the order of 1--10 eV \AA  we obtain a rough estimate for $v_c$ of
the order of 10$^5$--10$^6$ m/s. This leads to a temperature scale $T$
of 1--10 meV or 10--100 K. In the conservative estimate of 10 K
it is therefore conceivable that finite-size quantization effects pose no
limitations on the possible observation of TLL physics in local observables
such as the LDOS at the lowest experimentally obtained temperature of
order 5K.

Next, let us consider the higher dimensional coupling.
As a perturbation to the uncoupled wires, we consider the inter-chain hopping described by the Hamiltonian
\begin{multline}
\label{eq:4}
 \delta H = t_{\perp} \sum_{<i,j>,\sigma}\int dx \Bigl[\Psi^{\dag}_{i\sigma}(x)\Psi_{j\sigma}(x) +h.c.\Bigr] 
\end{multline}
For the repulsive interaction $K_c \approx 0.26$, we can neglect superconducting order caused by Cooper-pair hopping.
We do need to take density-density interaction and spin-exchange into account which,
if not present in the bare Hamiltonian, will be generated by second order processes from $\delta H_{t_{\perp}}$.
We can compactly write 
\begin{equation}
\label{eq:5}
\delta H= J_{\alpha} \sum_{<i,j>,\sigma}\int dx\, S_i^{\alpha}(x)S_j^{\alpha}(x)
\end{equation}
where $S_i^{\alpha}(x) =
\sum_{\sigma\sigma'}\Psi_{i\sigma}(x)\tau^{\alpha}_{\sigma\sigma'}
\Psi_{\sigma' i}(x)$, Here $\tau^0$ denotes the $2\times 2$ identity
matrix and $\tau^{1,2,3}$ the Pauli spin-matrices.
Assuming spin-rotation invariance, the RG equations to lowest order are \cite{GiamarchiBOOK,1995_Boies_PRL_74}
\begin{align}
\label{eq:6}
  \frac{d t_{\perp}}{dl} = \frac{6 - K_c - K_c^{-1}}{4}\\
  \frac{d J_{\alpha}}{dl}= \left(1 - K_c\right)J_{\alpha}+t_{\perp}^2.
\end{align}
Starting from small $t_{\perp}$ and $1/3<K_c<1$ we find that
$t_{\perp}$ grows quicker with the RG flow than $J_{\alpha}$,
initially, and thus one expects to find the transition temperature
$T_1$ for the dimensional crossover caused by $\delta H_{t_{\perp}}$
to occur before the temperature of spin or charge ordering caused by
$\delta H_{J_{\alpha}}$.
For $0<K_c<1/3$, $\delta H_{J_{\alpha}}$ always grows faster and hence $T_2$ is likely to occur first in all cases.
An estimate for $T_1$ may be obtained by neglecting the renormalization for $K_c$ and $J_{\alpha} $.
The dimensional crossover is then expected when the renormalized $t_{\perp}$ becomes comparable to the band width $t_{\parallel}$, which based on the
band-bottom energy of the Au/Ge(001) ARPES data we take to be of the
order of $100$ me V for discussion purposes.
The crossover energy is estimated as $T_1 \sim t_{\parallel}(t_{\perp}/t_{\parallel})^{\nu^{-1}}$, with $\nu =(6 - K_c - K_c^{-1})/6$, which gives
$T_1 \sim t_{\perp}$ for the non-interacting case $\nu = 1$.
Setting $T_1 = 10$ K and  $K_c = 0.26$ we obtain $t_{\perp} \sim 10$
meV as the maximal allowable inter-chain hopping, one tenth of the $t_{\parallel}$ value. 

Similar reasoning can in principle be applied to disorder, and estimates of the localization length $\xi_{\rm loc}$ can be obtained
\cite{GiamarchiBOOK,1988_Giamarchi_PRB_37} from which a temperature
follows by setting $L_T\sim \xi_{\rm loc}$. However, since there is no
reliable estimate for the disorder strength in Au/Ge(001), no quantitatively meaningful statement can be made here at present.\\

\subsection{Anomalous suppression of the DOS at the Fermi level}\label{app:dip}

\begin{figure}[h]
	\centering
	\includegraphics[width=0.5\textwidth]{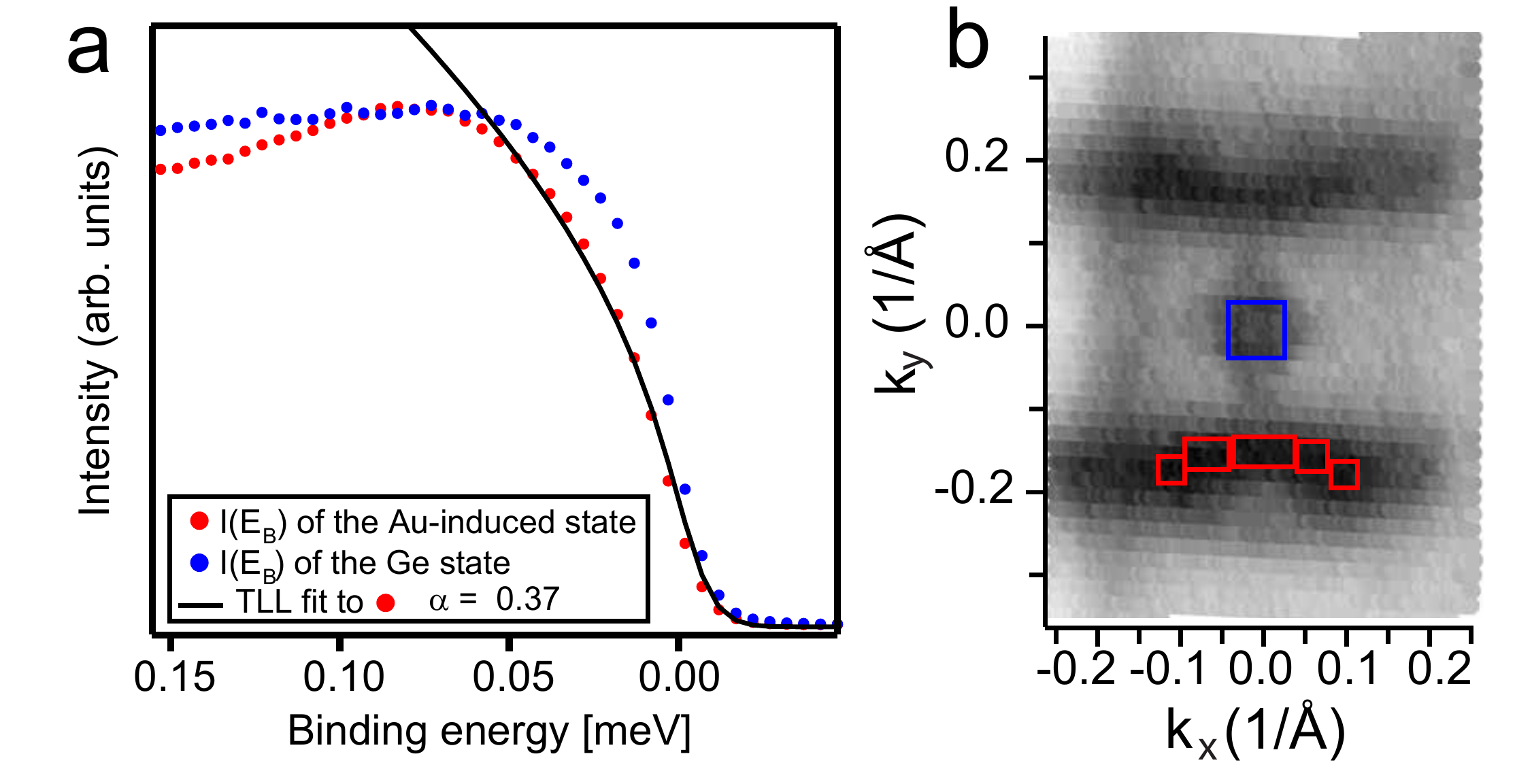}
	\caption{{\bf (a) k-integrated ARPES data yielding I(E) curves from the photoemission data presented in panel (b). The blue data-points corresponds to the I(E) curve for the bulk-related Ge state close to $\Gamma$, arrived at by integrating over the blue square in superimposed on the (b). The red curve shows the I(E) data from the Au-induced electron pocket, integrated over the red tiles shown in panel (b). The black line in panel (a) depicts a fit to the red I(E) curve from the Au-induced states based on the expectations for a TLL, based on Ref. \onlinecite{1993_Schoenhammer_JESRP_62}. Panel (b): I(k$_x$,k$_y$) map at E$_B$=15 meV. All ARPES data were measured with a photon energy of 21.2 eV at 20 K. }}
	\label{fig:zerobias}
\end{figure}

Here we present k-integrated I(E) curves from our ARPES measurements showing the incoherent nature of the states close to the Fermi level. In Fig. \ref{fig:zerobias}(a) two I(E) curves are presented, k-integrated over the bulk-related Ge state crossing the Fermi level at $\Gamma$ (shown in bue) and the Au-induced electron pocket state (shown in red). The k-integration cannot be particularly extensive, and the corresponding k-regions are highlighted with colour coded tiles in Fig. \ref{fig:zerobias}(b). While the I(E) trace derived from the bulk-related Ge states shows the clear Fermi-step charcterising the occupied states of a 2D metal, the I(E) trace from the Au-induced bands shows gradual suppression of the signal for energies approaching  E$_F$. This behaviour is similar to that reported in Refs. \cite{meyer11prb,blumenstein13jpcm,meyer14prb,blumenstein11natphy}.
The black solid line in panel (a) shows a fit to the red (I(E) curve using the formula for the density of occupied states based on Ref. \onlinecite{1993_Schoenhammer_JESRP_62}:
\begin{equation}
\rho^{h}_{x} \left( E,T\right) \propto T^{\alpha} e^{-\frac{E}{2k_{B}T}} \left| \Gamma \left( \frac{1+\alpha}{2} +i \frac{E}{2\pi k_{B}T} \right) \right|^2 
\end{equation}
in which E the binding energy, T the temperature and k$_B$ the
Boltzmann constant. As can be seen from the figure, an acceptable fit
of the data can be arrived at by taking an alpha of 0.37.

\end{document}